\documentclass[
    ,final            
  ]
  {aipproc}

\layoutstyle{6x9}

\begin{document}
 
\title{Absorption  Features in Spectra of Magnetized Neutron Stars} 
 
 
\author{V. Suleimanov}{ 
  address={Insitute for Astronomy and Astrophysics, Kepler Center for Astro and Particle Physics, Eberhard Karls University, Sand 1, 72076 T\"ubingen, Germany}, 
altaddress={Kazan Federal University, Kremlevskaya str. 18, 42008 Kazan, Russia}, 
email={suleimanov@astro.uni-tuebingen.de} 
} 

\author{V. Hambaryan}{ 
  address={Astrophysikalisches Institut und Universit\"ats-Sternwarte Jena, Schillerg\"asschen 2-3, 07745 Jena, 
 Germany} 
} 

\author{A.~Y. Potekhin}{ 
  address={Ioffe Physical-Technical Institute,  Politekhnicheskaya str., 26, St. 
Petersburg 194021, Russia} 
} 

\author{G.~G. Pavlov}{ 
  address={Pennsylvania State University, 525 Davey Lab., University Park, PA 16802, USA} 
} 

\author{M.~van~Adelsberg}{ 
  address={Kavli Institute for Theoretical Physics, Kohn Hall, University of California, Santa Barbara, CA 93106, USA} 
} 

\author{R.~Neuh\"auser}{ 
  address={Astrophysikalisches Institut und Universit\"ats-Sternwarte Jena, Schillerg\"asschen 2-3, 07745 Jena, 
 Germany} 
} 

\author{K. Werner}{ 
  address={Insitute for Astronomy and Astrophysics, Kepler Center for Astro and Particle Physics, Eberhard Karls University, Sand 1, 72076 T\"ubingen, Germany}, 
} 
 
\begin{abstract} 
The X-ray spectra of some magnetized isolated neutron stars (NSs) show 
absorption features 
with equivalent widths  (EWs) of 50 -- 200 eV, whose  
 nature is not yet well known. 
 
To explain the prominent absorption features in the soft X-ray spectra  
of the highly magnetized ($B \sim 10^{14}$ G) X-ray dim isolated NSs (XDINSs), 
we theoretically investigate different NS local surface models, including naked condensed iron 
surfaces and partially ionized hydrogen model atmospheres, 
with semi-infinite and thin atmospheres above the 
condensed surface. We also developed a code for computing light curves and integral emergent 
spectra of magnetized neutron stars with various temperature and magnetic field distributions over the NS surface. 
We compare the general properties of the computed and observed light curves and integral spectra for XDINS RBS\,1223   
and conclude that the observations can be explained by a thin hydrogen atmosphere above the condensed iron 
surface, while the presence of a strong toroidal magnetic field component 
on the XDINS surface is unlikely. 
 
We suggest that the harmonically spaced absorption 
features in the soft X-ray spectrum of the central compact object (CCO) 1E\,1207.4--5209 (hereafter 1E\,1207) 
correspond to peaks  
in the energy dependence of the free-free opacity in a quantizing magnetic field, known as quantum oscillations. 
To explore observable properties of these quantum oscillations, 
we calculate models of 
hydrogen NS atmospheres 
with $B \sim 10^{10}$--$10^{11}$ G (i.e., electron cyclotron energy 
$E_{c,e}\sim 0.1$--1 keV) and $T_{\rm eff} 
 = 1$--3 MK. Such conditions are 
thought to be typical for  1E\,1207. 
We show that observable features at the electron cyclotron harmonics with EWs $\approx$ 100 -- 200 eV 
can arise due to these quantum oscillations. 
 
\end{abstract} 
 
\maketitle 
 
 
\section{Absorption feature in the spectrum of RBS\,1223}

XDINSs are a new class of pulsing X-ray sources with soft ($T_{\rm eff} \sim 10^6$ K) thermal-like spectra  
(see \cite{Haberl:07} for a review). Here we consider the most extreme member of the XDINSs -- RBS\,1223, which is  pulsing with an amplitude 
of 18\% \cite{Swopeetal:05} and shows  
 the most prominent absorption feature with an equivalent width (EW) of $\approx 200$ eV at $\approx 0.3$ keV \cite{Swopeetal:07}. 
We investigate what kind of radiating surface can qualitatively explain the observed properties of this XDINS.  
For this aim, three local models are considered:  naked condensed surfaces, semi-infinite magnetized model atmospheres, and thin 
magnetized model atmospheres above condensed surfaces.

\begin{figure} 
  \includegraphics[height=0.23\textheight]{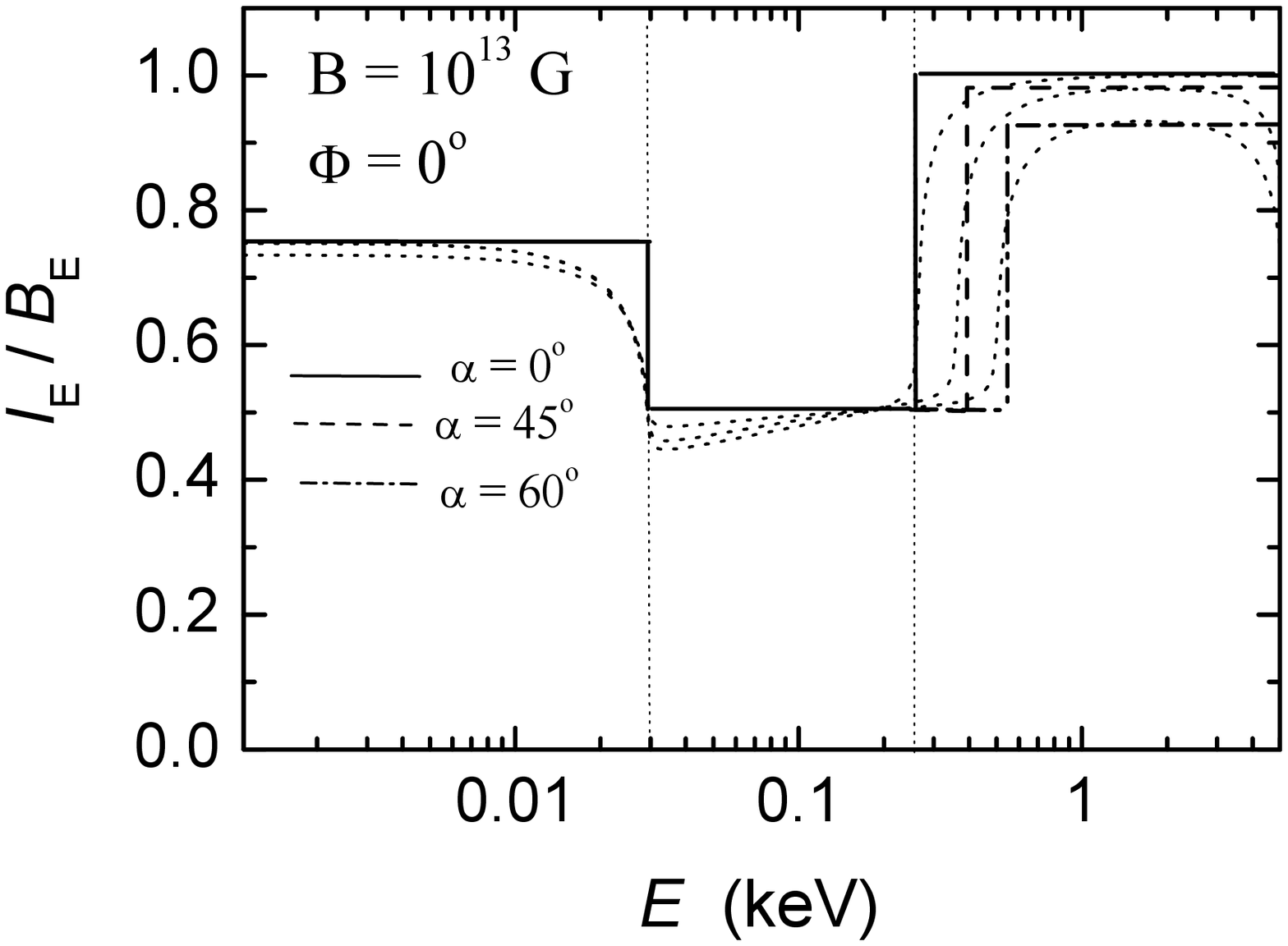} 
  \includegraphics[height=0.23\textheight]{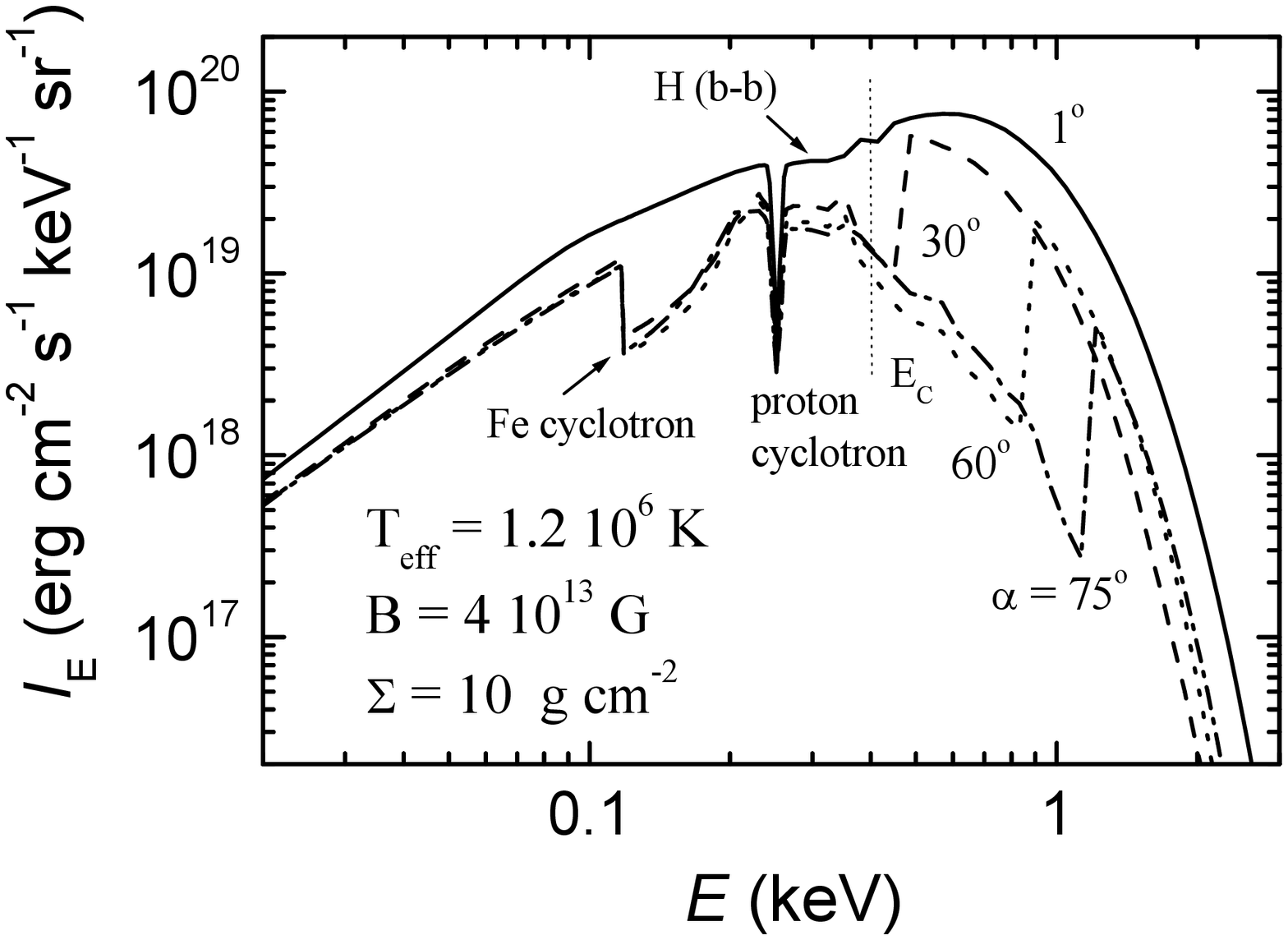} 
  \caption{ 
{\it Left:}Approximation of the dimensionless emissivity as a function of photon energy $E$ for the case of 
a condensed iron surface at $B = 10^{13}$ G. The magnetic field is normal to the surface. 
 {\it Right:} Emergent specific intensity spectra of a model atmosphere above a condensed iron surface.
} 
\end{figure}

Spectra of naked condensed iron surfaces \cite{vA:05} are  represented using a simple  analytical approximation (see Fig.\,1, left panel). 
In the free-ion approximation,  a broad absorption feature exists in the spectrum. It 
lies between the ion cyclotron energy $E_{\rm c,Fe} \approx 0.32~ (B / 10^{14} G)$ keV, and 
some boundary energy $E_{\rm C}$, which weakly depends on the magnetic field and strongly depends on the angle $\alpha$  
between the magnetic field and the photon propagation direction ($E_{\rm C} \approx~ 0.55$ keV at $B\approx 10^{14}$ G). 
 
Some examples of magnetized semi-infinite partially ionized hydrogen model atmospheres were computed by us in \cite{SPW:09}. 
Here we 
computed  thin (with surface densities $\Sigma \approx$ 1 -- 10 g cm$^{-2}$)  model atmospheres above condensed iron surfaces 
using their radiation properties as inner boundary conditions (Fig.\,1, right panel).  
Because of the condensed surface emission properties, 
 broad absorption features appear between  $E_{\rm c,Fe}$ and  $E_{\rm C}$, together with a strong edge  
in the angular distribution of the radiation 
at an energy between $E_{\rm C}$ and $4E_{\rm C}$ and corresponding edges in the specific intensity spectra. The absorption feature's  
EW may reach 300 -- 400 eV in these models, whereas EWs are smaller than 100 eV for the absorption features in the spectra 
 of the semi-infinite atmospheres. The emergent spectra of the semi-infinite  
and thin model atmospheres can be represented by diluted  
black-body spectra with one or two Gaussian absorption lines. The angular  
distribution of the radiation is represented by a simple step function 
between $E_{\rm C}$ and $4E_{\rm C}$ and is assumed to be isotropic for other photon energies.  
 
To calculate the XDINS spectra and light curves, we use the model of a slowly rotating  
spherical isolated neutron star with a given compactness $M/R$, 
with account of the gravitational redshift 
and light bending.  The 
distributions of the local color temperatures and magnetic field strengths over the stellar surface  
\begin{equation} 
\label{um4} 
  T^4 = T_{\rm p}^4 \frac{\cos^2\theta}{\cos^2\theta + a\,\sin^2\theta } + T_{\rm min}^4, 
~~~~B = B_{\rm p} \sqrt{\cos^2\theta + a\,\sin^2\theta} 
\end{equation}     
were taken from \cite{PerAzetal:06}. 
Here $\theta$ is the magnetic colatitude, $T_{\rm p}$ and $B_{\rm p}$ are the temperature and the magnetic field strength at the poles, 
and the parameter $a$ characterizes the geometry of the magnetic field: $a=0.25$ corresponds to the pure dipole field, and $a\gg 1$ corresponds  
to a strong toroidal component (in this case $a^2 \approx B_{\rm tor}/B_{\rm p}$). 
 
 We consider a neutron star model with $kT_{\rm p} = 
0.15$ keV, $B_{\rm p} = 6 \times 10^{13}$ G, and $z = 0.2$. Three temperature and magnetic field 
distributions are used: $a=0.25$, $a=60$ and two uniform bright spots with angular radii 
$\theta_{\rm sp} = 5^{\circ}$ and the dipole magnetic field. 
 Local 
spectra of the three models of NS radiating surfaces are presented by analytical functions (see above). 
The central energies of the local absorption features depend on the local magnetic field. 
 
\begin{figure} 
  \includegraphics[height=0.23\textheight]{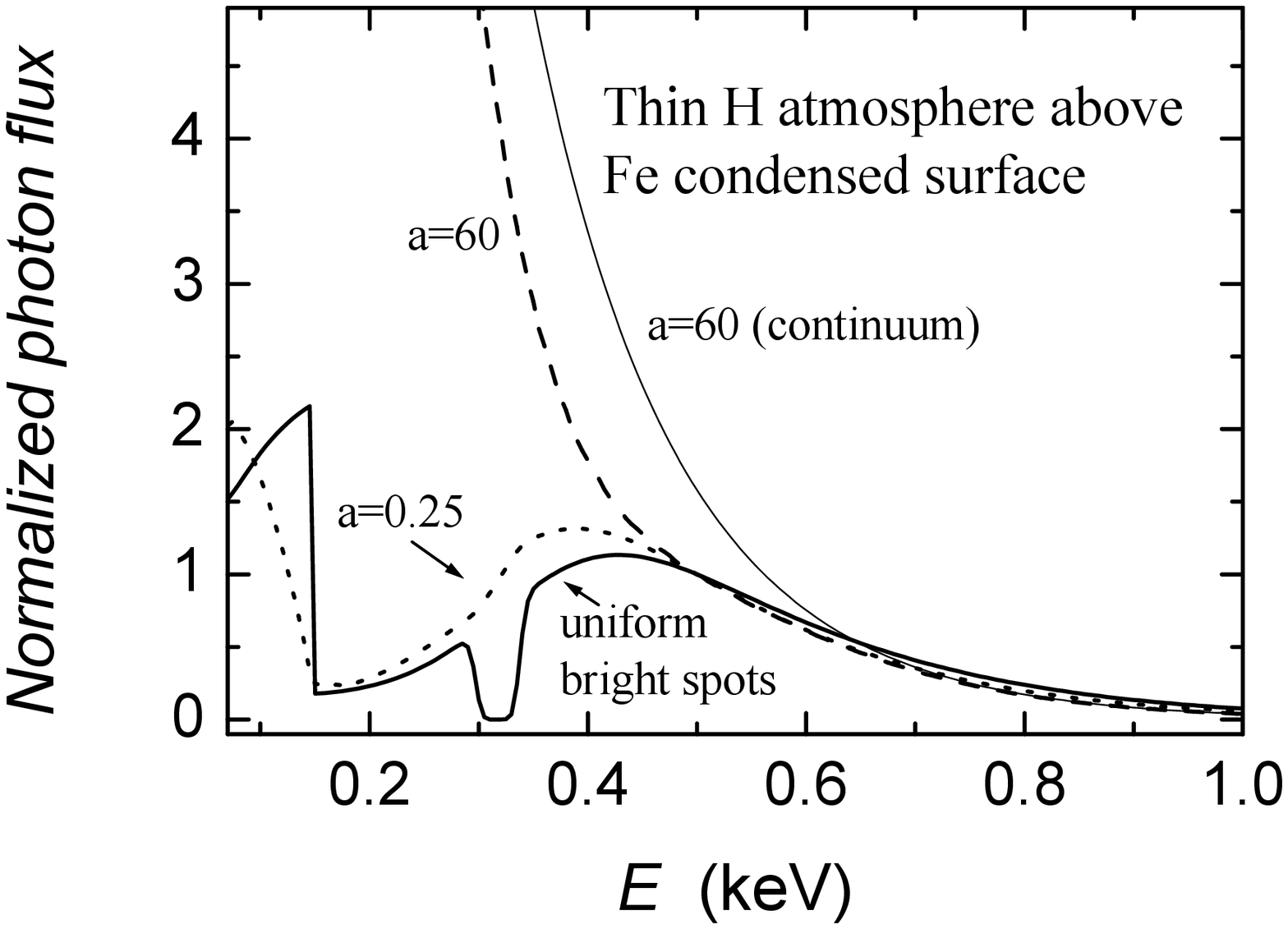} 
  \includegraphics[height=0.23\textheight]{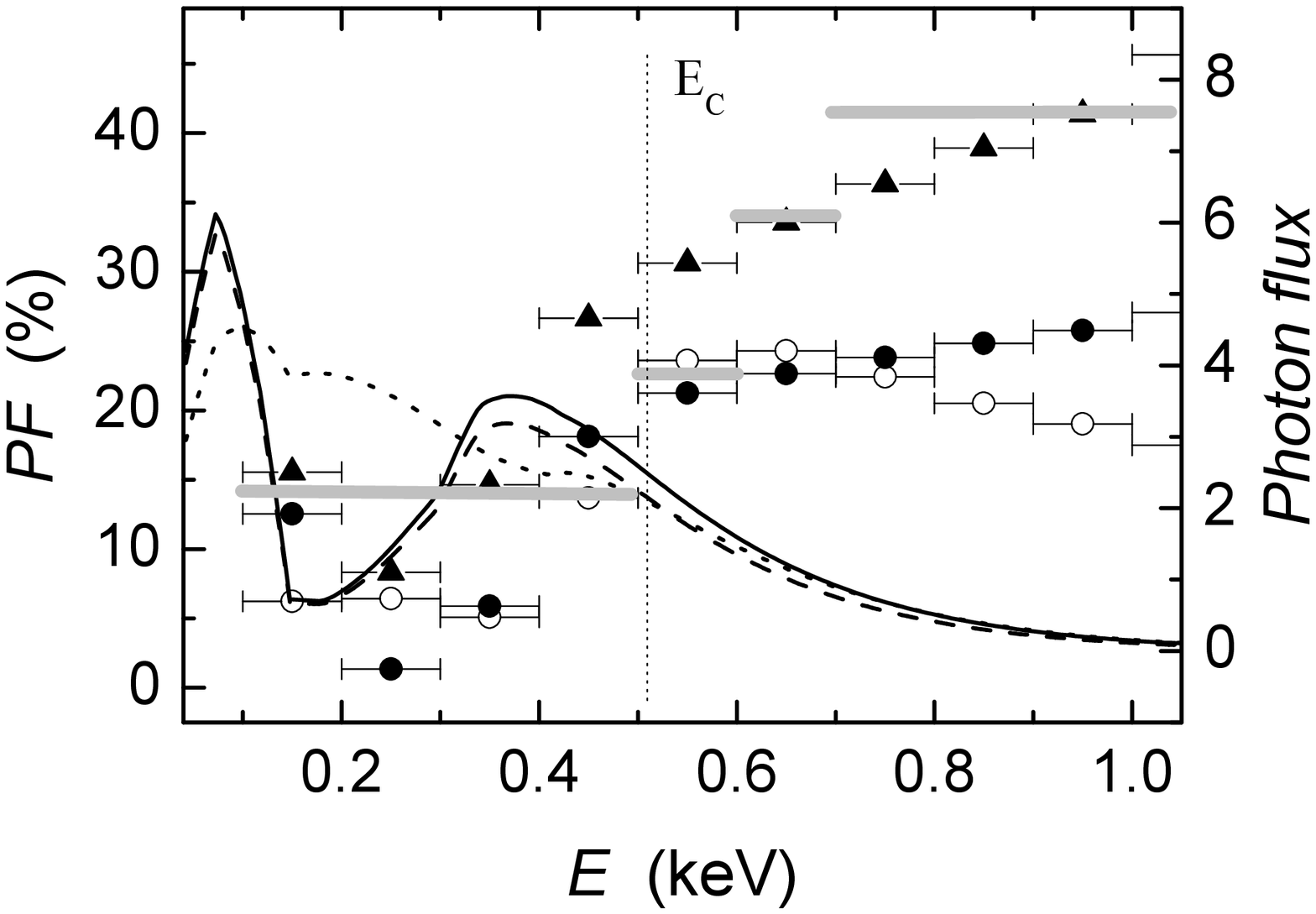} 
  \caption{ 
{\it Left:} Phase averaged photon spectra of the neutron star models. 
The local spectra are isotropic blackbodies with absorption features as in the spectra of the thin hydrogen 
model atmosphere above the condensed iron surface.  
 {\it Right:} Energy dependence of pulsed fractions (symbols and left vertical 
axis) together with averaged photon spectra (lines and right 
vertical axis) for the neutron star 
models with $a$=0.25. 
Open circles and dotted curve - a naked condensed iron surface. 
Filled circles and solid curve - a thin atmosphere above condensed iron surface spectra with corresponding angular distribution. 
Triangles and dashed curve - the same local spectra, but with slightly different pole temperatures: 
$T_{\rm p,1}$ = 0.15 keV, $T_{\rm p,2}$ = 0.14 keV. The observed RBS\,1223 pulsed fractions  
are shown by thick gray horizontal lines. 
} 
\end{figure}

We  demonstrated that the EWs of the absorption features in the integral NS 
spectra are similar to the local EWs on the neutron star surfaces.  
Therefore, the EW of the absorption line in the spectrum of RBS\,1223 (about 200 eV) can only be explained by a thin atmosphere above the condensed 
iron surface. The same local model with a smooth temperature distribution over the neutron star 
surface ($a$\,=\,0.25) and slightly different pole temperatures can 
provide the pulsed fraction observed  in RBS\,1223 (see Fig.\,2). A strong toroidal magnetic field component  on the XDINS 
surfaces ($a \gg$ 1) seems unlikely because 
 such models exhibit a too wide, smoothed 
absorption feature. 
See details in \cite{SH:10}. 
 
\section{Absorption features in the spectrum of 1E\,1207} 
  
The absorption features in the spectrum of 1E\,1207, 
centered at about 0.7 and 1.4 keV,  
were discovered in \cite{Sanwal:02}.  
We demonstrate that these lines can  
arise due to quantum resonances in the free-free absorption of photons  
in a magnetic field  at energies of electron cyclotron line and its harmonics \cite{pp:76}. 
 
We present 
computations of fully ionized hydrogen atmospheres of  
NSs for magnetic fields $B \sim 10^{10}$--$10^{11}$ G and the effective temperatures 1 -- 3 MK (see Fig.\,3). 
In these models the electron cyclotron energy
is within the observed range of energies.  
 
 The electron quantum oscillations are best observable at moderately large values of the 
quantization parameter, 
$0.5 \le b_{\rm eff}=E_{\rm c,e}/kT_{\rm eff} 
\le 20$, 
when the quantization is significant but the features are not too far in the Wien tail 
of the spectrum. 
The equivalent widths of the absorption features 
reach $\sim 100$--200 eV in the examples considered; they grow with increasing $b_{\rm eff}$ 
and are lower for higher harmonics. 
Therefore, the observed features can be interpreted as 
 caused by the quantum oscillations. See details in \cite{SPW:10}.

\begin{figure} 
  \includegraphics[height=0.23\textheight]{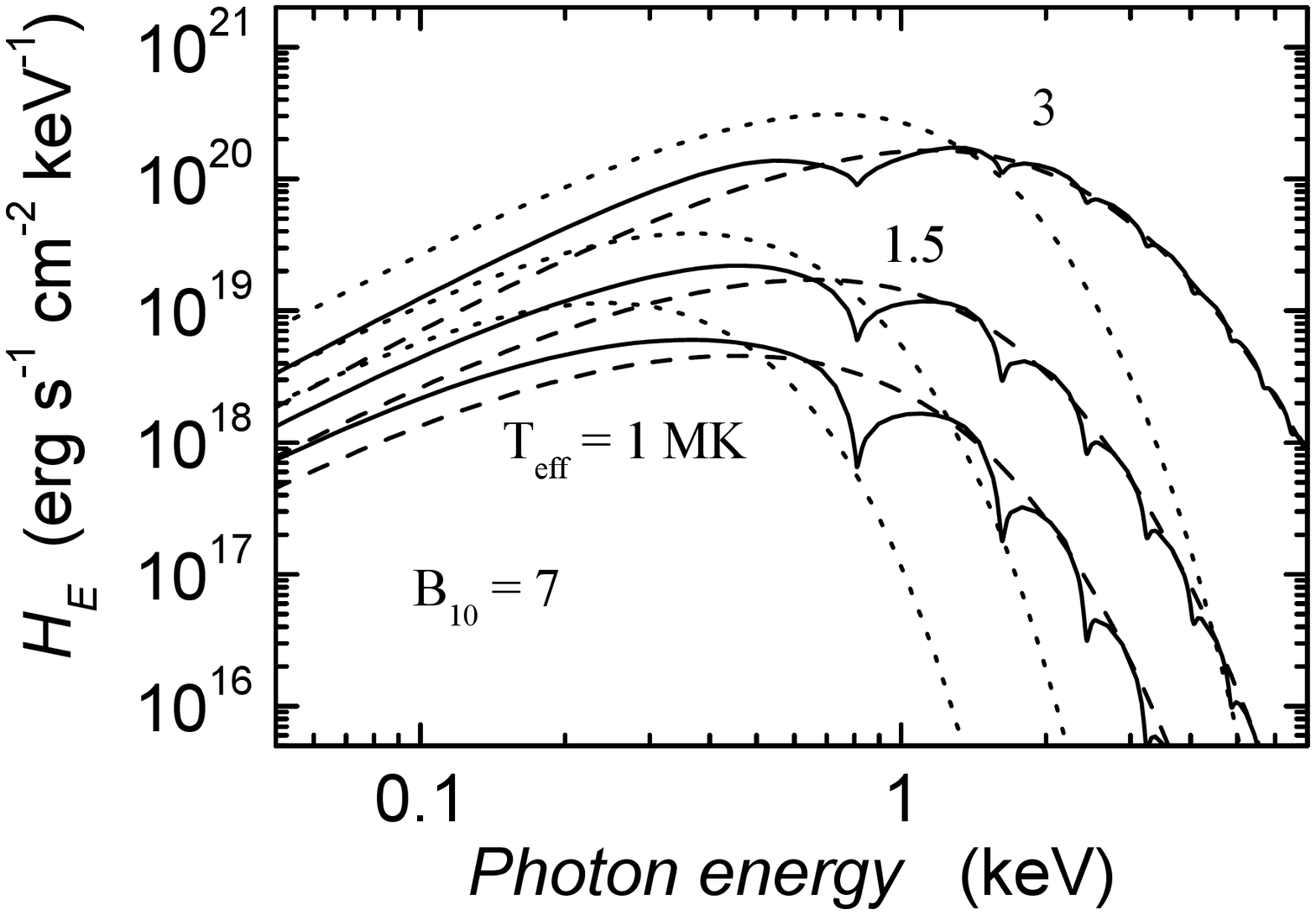} 
  \includegraphics[height=0.23\textheight]{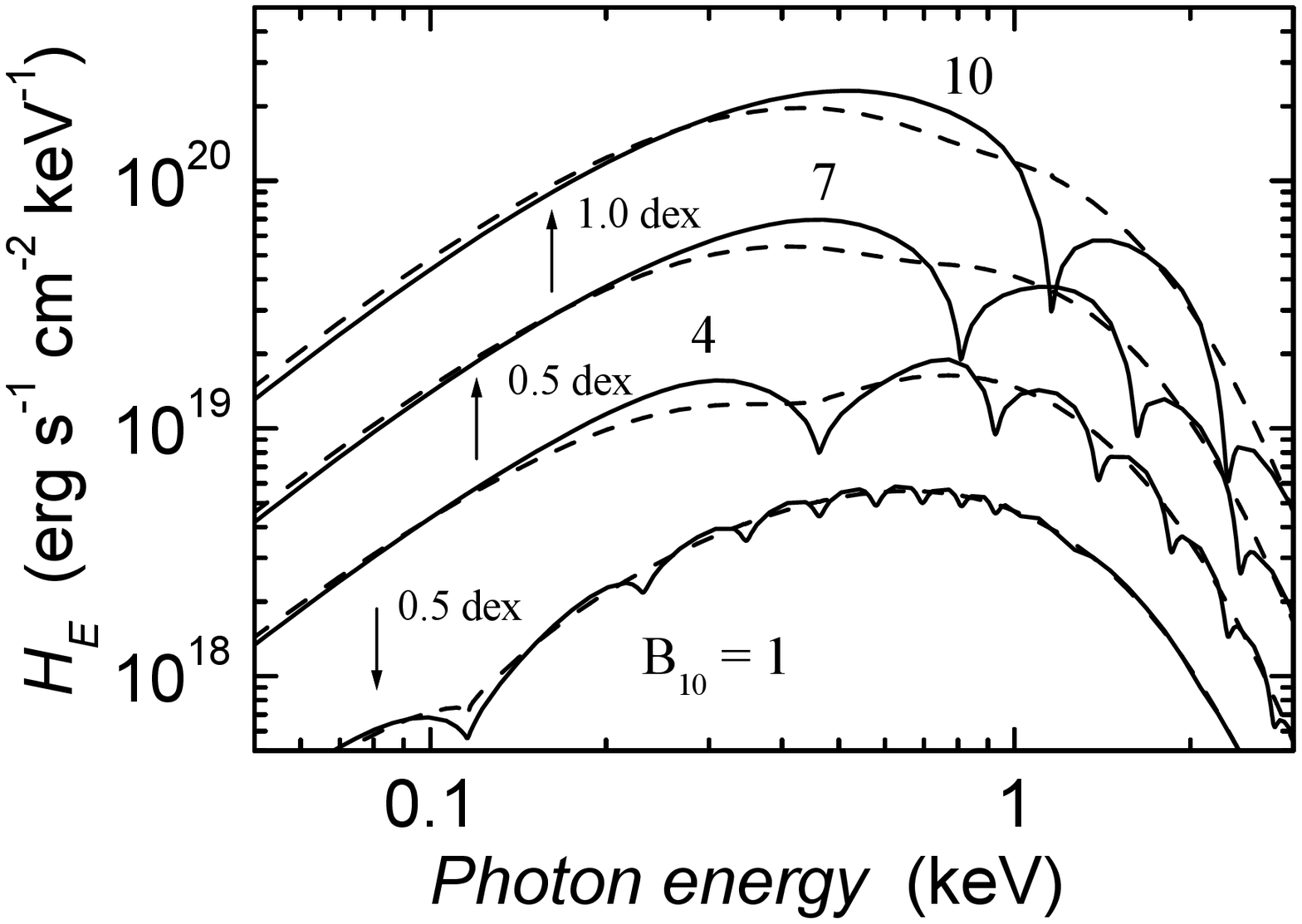} 
  \caption{ 
{\it Left:} Emergent spectra for  
NS atmospheres with  
 magnetic field $B=7\times 10^{10}$ G (solid curves) and $B=0$ (dashed 
curves) for three 
effective temperatures. 
The dotted curves are blackbody spectra for the same temperatures. 
 {\it Right:} Emergent spectra  
for magnetic  
NS atmospheres with $T_{\rm eff}=1.5$ MK and 
four different magnetic fields,  
calculated with magnetic and non-magnetic Gaunt factors. 
} 
\end{figure}

\begin{theacknowledgments} 
  
This work is supported by the DFG grant SFB / Transregio 7 ``Gravitational Wave Astronomy'' (V.S., V.H.),  
Russian Foundation for Basic Research (grant  08-02-00837, A.P.), Rosnauka (grant NSh-3769.2010.2, A.P.), and 
NASA grant NNX09AC84G (G.P.). 
\end{theacknowledgments} 

\bibliography{Suleimanov_Valery2}
\bibliographystyle{aipproc}

\end{document}